\documentclass[a4paper,11pt]{article}
\pdfoutput=1 %

\usepackage{jcap} %
\usepackage{aas_macros}
\usepackage{newtxtext,newtxmath}
\usepackage{multirow}
\usepackage[T1]{fontenc}
\usepackage{ae,aecompl}
\usepackage{graphicx}	%
\usepackage{amsmath}	%
\usepackage{amssymb}	%
\usepackage{multicol}	%
\usepackage{bm}		%
\usepackage{pdflscape}	%
\usepackage{hyperref}
\usepackage{ulem} %
\usepackage{lineno} %
\usepackage{lipsum}
\usepackage{mathtools}

\usepackage{graphicx}	%
\usepackage{amsmath}	%
\usepackage{amssymb}	%

\newcommand{\RNum}[1]{\uppercase\expandafter{\romannumeral #1\relax}}

\usepackage[dvipsnames]{xcolor} %

\newcommand{\redmapper}{redMaPPer}

\newcommand{\Mpc}{\rm{Mpc}}

\newcommand{\hMpc}{h^{-1} \Mpc}

\newcommand{\hinv}{h^{-1}}

\newcommand{\Omegam}{\Omega_{{\rm m}}}
\newcommand{\Omegab}{\Omega_{{\rm b}}}

\title{\boldmath Deciphering baryonic feedback with galaxy clusters}

\author[a,b,c,1]{Chun-Hao To \note{Corresponding author.}}
\author[d]{Shivam Pandey}
\author[e,f]{Elisabeth Krause}
\author[a,b]{Nihar Dalal}
\author[g,h]{Dhayaa Anbajagane}
\author[a,c]{David H. Weinberg}
\emailAdd{to.87@osu.edu}
\emailAdd{sp4204@columbia.edu}
\emailAdd{krausee@arizona.edu}
\emailAdd{dalal.64@osu.edu}
\emailAdd{dhayaa@uchicago.edu}
\emailAdd{weinberg.21@osu.edu}

\affiliation[a]{ Center for Cosmology and AstroParticle Physics (CCAPP), Ohio State University, Columbus, OH 43210, USA}

\affiliation[b]{  Department of Physics, Ohio State University, Columbus, OH 43210, USA}
\affiliation[c]{  Department of Astronomy, Ohio State University, Columbus, OH 43210, USA}
\affiliation[d]{Department of Physics, Columbia University, New York, NY, USA 10027}
\affiliation[e]{  Department of Physics, University of Arizona, Tucson, AZ 85721, USA}
\affiliation[f]{  Department of Astronomy/Steward Observatory, University of Arizona, 933 North Cherry Avenue, Tucson, AZ 85721-0065, USA}
\affiliation[g]{Department of Astronomy and Astrophysics, University of Chicago, Chicago, IL 60637, USA}
\affiliation[h]{Kavli Institute for Cosmological Physics, University of Chicago, Chicago, IL 60637, USA}

\abstract{
Upcoming cosmic shear analyses will precisely measure the cosmic matter distribution at low redshifts. At these redshifts, the matter distribution is affected by galaxy formation physics, primarily baryonic feedback from star formation and active galactic nuclei. Employing measurements from the \textsc{Magneticum} and \textsc{IllustrisTNG} simulations and a dark matter + baryon (DMB) halo model, this paper demonstrates that Sunyaev-Zel’dovich (SZ) effect observations of galaxy clusters, whose masses have been calibrated using weak gravitational lensing, can constrain the baryonic impact on cosmic shear with statistical and systematic errors subdominant to the measurement errors of DES-Y3 and LSST-Y1. We further dissect the contributions from different scales and halos with different masses to cosmic shear, highlighting the dominant role of SZ clusters at scales critical for cosmic shear analyses. These findings suggest a promising avenue for future joint analyses of Cosmic Microwave Background (CMB) and lensing surveys.
}
\begin{document}
\maketitle
\flushbottom

\section{Introduction}
Weak gravitational lensing is a powerful probe of cosmic structure formation that measures the total (dark and luminous) matter distribution without requiring a bias model relating matter to more directly observable astrophysical tracers. While the total matter distribution is shaped primarily by gravity, it is also affected by small-scale astrophysical processes, particularly baryonic feedback from galaxy formation. Hence, a correct cosmological interpretation of weak lensing measurements has to accurately account for the impacts of galaxy formation physics on the total  matter distribution \citep[see, e.g.,][ for a review]{Chisari2019}. This is a challenging modeling problem: much of the weak lensing signal-to-noise is derived from scales where gravitational evolution is already non-linear, with the physics of galaxy formation adding further non-linearities, limiting the use of perturbative approaches. Hence, modeling galaxy formation’s impact on the total (non-linear) matter distribution relies on numerical simulations.

Simulations that incorporate galaxy formation physics in addition to gravity can, in principle, predict the modulation of matter clustering due to feedback from galaxy formation physics \citep[see, e.g.,][for a review]{2020NatRP...2...42V}. Practically, however, such simulations require modeling physics across a wide range of scales, making them notoriously computationally expensive. Furthermore, galaxy formation incorporates a wide range of uncertain physical processes that cannot be simulated ab initio, making it hard to judiciously choose physical processes to be modeled and free parameters associated with them. %
Recent efforts have produced simulations spanning a wide range of free parameters regarding the physics of baryonic feedback from galaxy formation, with the aim of marginalizing over this uncertain galaxy formation physics in cosmological analyses \citep{2021ApJ...915...71V,2023MNRAS.523.2247S,2023MNRAS.526.4978S}. %
However, questions remain about whether the range of galaxy formation physics explored in simulations is wide enough to model the range of possible impacts of baryonic physics in the real universe. Conversely, uncertainties in galaxy formation models and the flexibility of the models used to summarize the simulation results limit the precision of the cosmological interpretation of weak gravitation lensing.

Many approaches have been developed to mitigate this uncertainty of baryonic feedback from galaxy formation. The key cosmology analyses of the Year 1 (Y1) and Year 3 (Y3) Dark Energy Survey (DES) data sets \citep{Y3_Secco, Y3_Amon, DESY3, DESY1}  remove cosmic shear measurements on scales expected to be affected by baryonic physics. While this approach is the most conservative, it removes about one-third of the signal-to-noise. \cite{Huang2021} reanalyze DES-Y1 data, including cosmic shear on scales affected by baryonic physics, and marginalize uncertainties of baryonic physics predicted by $11$ numerical simulations that implement different galaxy formation models. 
While the conclusion depends on the ranges of baryonic physics explored (see also \citep{2023arXiv231108047X}), they find that the uncertainties of baryonic physics are sufficiently large to remove most of the constraining power of the additional data in the most conservative setting. Further, \cite{chen2023, Arico2023, 2022MNRAS.514.3802S} incorporate baryon correction models that alter gravity-only simulations according to a dark matter + baryon (DMB) halo model to mimic the impact of baryonic physics on matter clustering. They reanalyze survey datasets and find similar conclusions as \cite{Huang2021}, that most of the signal-to-noise on additional cosmic shear measurements relative to key cosmological analyses goes to constrain baryonic physics instead of cosmology.

Adding more observables to constrain baryonic physics is, therefore, important to convert the additional signal-to-noise in small-scale cosmic shear measurements to cosmological constraints. %
Cosmic gas is redistributed by baryonic physics that modulates the total matter distribution, making observations of its distribution particularly constraining on relevant baryonic physics \citep{vandaalen2011}.
Numerous programs \citep{Troster2021,Schneider22, 2022JCAP...04..046N, Ferreira23, Grandis2023,2024arXiv240103510Z, ACTxDEStsZ2, ACTxDEStsZ1} have explored observations of the cosmic gas distribution to constrain modulations of matter clustering due to baryonic feedback, some of which \citep{Troster2021,Schneider22} have been combined with weak gravitational lensing analyses, leading to different levels of improvements on cosmological constraints. %
Specifically, \citep{Troster2021} employ the cross-correlation of thermal Sunyaev-Zel'dovich (tSZ) signal detected in the cosmic microwave background (CMB) and weak lensing shear. \citep{Schneider22} explore the cross-correlation of kinematic Sunyaev-Zel'dovich (kSZ) signal detected in the CMB, weak lensing shear, and spectroscopically selected galaxies. \citep{Ferreira23} use X-ray background and cosmic shear cross-correlation. \citep{Grandis2023} and \citep{Schneider22} explore the measured gas contents from X-ray observations of galaxy clusters. While promising, interpretations of these observations face several theoretical challenges. Constraining the modulation of the matter distribution from these observations is a two-step process: one has to link observations to the cosmic gas distribution and then link the cosmic gas distribution to the total matter distribution. Each of these steps incurs additional modeling complexities. While the second step is shared by all cosmic gas probes, the first step varies with each observational method. For instance, kSZ observations measure the electron distribution around galaxies, whose interpretation requires a model to connect those galaxies with their associated dark matter halos \citep[see, e.g.,][ for a review]{2018ARA&A..56..435W}. All sky X-ray background measurements can contain unresolved point sources such as active galactic nuclei and low-mass galaxy clusters \citep{1990ARA&A..28..657M, 1998MNRAS.295..641M, 1998A&A...329..482H, 2003A&A...399....9P,2001ApJ...560..544W, 2004MNRAS.352L..28W}. Finally, tSZ measures the electron pressure, whose connections to total thermal pressure require assumptions that the electrons and protons are in thermal equilibrium. This assumption is questionable at the outskirts of the halos  \citep{2009ApJ...701L..16R,2015ApJ...808..176A,2015A&A...579A..13V, 2022MNRAS.514.1645A, 2024MNRAS.527.9378A} and has not been validated in cosmological hydrodynamic simulations.

In this paper, we demonstrate the prospect of using the mean mass of tSZ galaxy clusters in bins of their tSZ signal, $\langle M|Y_{500c}\rangle$, to constrain the modulation of matter clustering due to baryonic feedback\footnote{\citep{2023MNRAS.525.1779P} has considered the prospect of using the $\langle Y|M\rangle$ of low mass halos to constrain the modulation of matter power spectra. But, we note that those studies are dominated by halos with a mass lower than our mass limit.}. The link between measurements of cosmic baryonic physics in the form of cluster mean mass--tSZ scaling relations and the modulation of the total matter distribution is established via a DMB halo model \citep{methodpaper}. This approach offers unique advantages that complement the existing constraints from X-ray cluster gas fraction measurements \citep{Grandis2023}. For instance, while tSZ clusters typically have higher mass limits compared to X-ray clusters, their selection function is less redshift dependent \citep[e.g.][]{Weinberg2013},  making them well-suited for cosmic shear analyses, which are sensitive to the matter distribution across a wide range of redshifts. In addition, mass estimates of individual clusters are noisy due to measurement and modeling uncertainties. This is especially true for the weak gravitational lensing mass calibration, where the signal is small. Stacking of many clusters to boost the signal-to-noise is thus common in ongoing weak lensing studies \citep{SPTDES1,SPT2}, making them only sensitive to the mean mass--observable relation of the considered samples \citep[see, e.g., section 6.3.3 of][]{Weinberg2013}. Therefore, the mean mass--tSZ signal relation $\langle M|Y_{500c}\rangle$ considered in this paper is directly connected to the data, requiring fewer assumptions on the observable--mass relation than the gas fraction-cluster mass relation $\langle f_{\rm gas}|M\rangle$. 

The $\langle M|Y_{500c}\rangle$ measured from SZ clusters only constrains baryonic redistribution in high mass halos. A critical insight, validated in this paper using analytic halo models and hydrodynamic simulations, is that these high mass halos contribute significantly to the impact of baryons on the cosmic shear correlation function so that constraints from $\langle M|Y_{500c}\rangle$ are enough to sharply reduce the baryonic physics uncertainties in cosmic shear analyses. This insight is aligned with the findings in \citep{2023MNRAS.523.2247S}, where the authors show that baryonic processes in high-mass halos are connected to the global matter power spectrum suppression. We anticipate that this approach can also mitigate uncertainties in the interpretation of galaxy-galaxy lensing measurements, but we defer the investigation of this topic to future work.

The paper is organized as follows. In section \ref{sec:highlevel}, we briefly summarize the main result of the paper by demonstrating the proposed analyses on two numerical simulations. In section \ref{sec:theory}, we detail the modeling framework. Section \ref{scalemass} demonstrates the significance of galaxy clusters in constraining the modulation of matter clustering due to baryonic feedback. Specifically, we dissect the sensitivity of cosmic shear to different physical scales and halo mass ranges. In section \ref{sec:Outlook}, we discuss the implications of the proposed analyses on cosmic shear studies in DES-Y3, LSST-Y1, LSST-Y6, and Roman.  
We conclude in section \ref{sec:conc}. In appendix \ref{app:hydro}, we compare the modulation of matter clustering and power spectra predicted by different hydrodynamic simulations and three empirical models. Throughout the paper, we adopt the following notation for halo mass:  $M_{200, 500c}=\frac{4}{3}\pi r_{200, 500 c}^3 [200, 500]\rho_{\rm c}$, and $r_{200, 500c}$ is the radius at which the averaged enclosed density of the halo is $200, 500$ times larger than the critical density of the universe $\rho_{\rm c}$. In the calculations associated with $M_{200, 500c}$, we assume an NFW profile.  %

\begin{figure*}
\centering
    \includegraphics[width =0.95\textwidth]{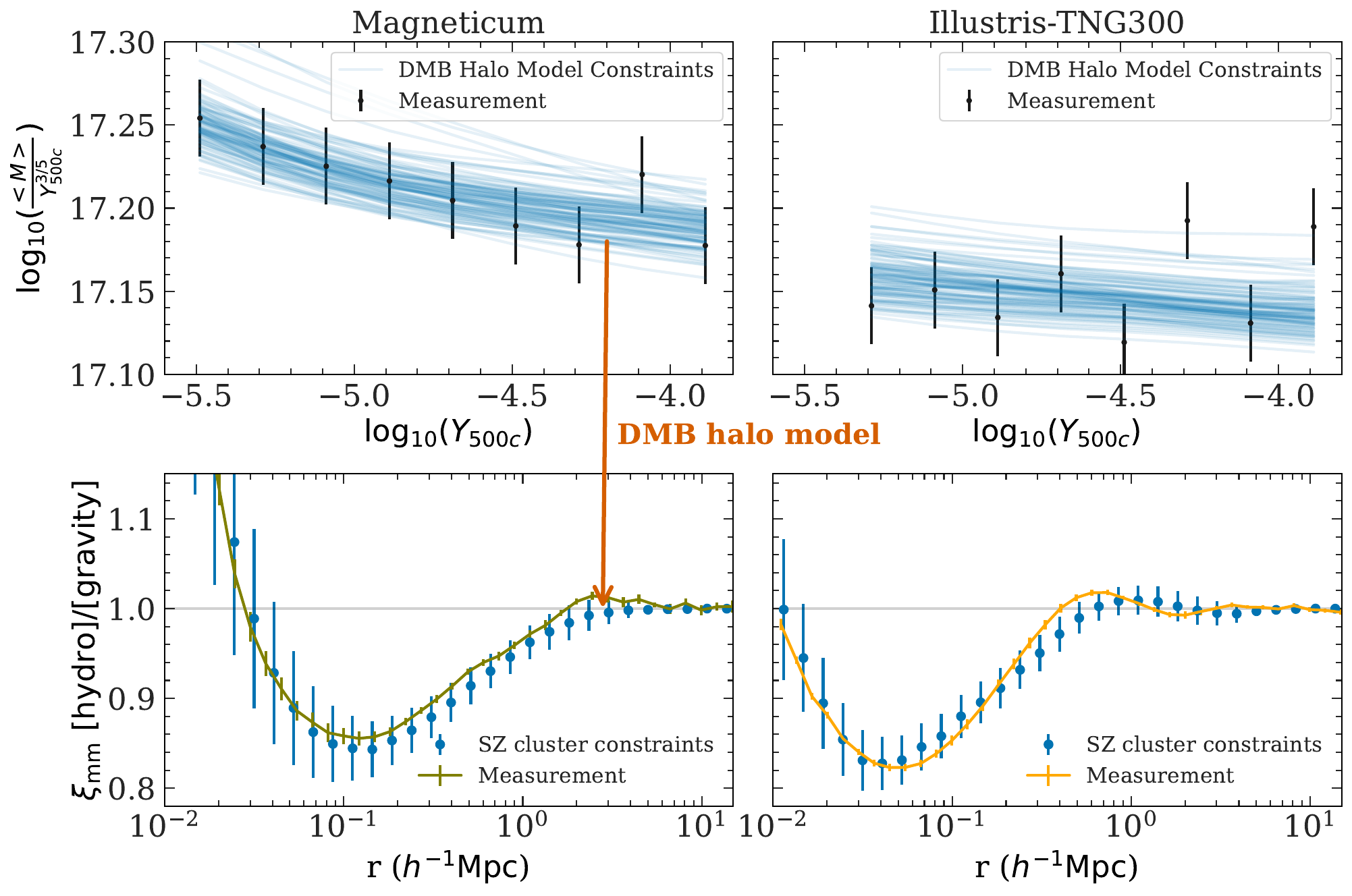}
    \caption{Validation of our approach to constrain modulations of matter correlations using galaxy cluster observables on two different hydrodynamical simulations, \textsc{Magneticum} (left) and \textsc{IllustrisTNG} (right). We use measurements of the cluster tSZ scaling relation $\langle M|Y_{500c} \rangle$ measured in these simulations, shown by the \textit{black symbols} on the top panels, to constrain parameters of the halo model, for which we show a hundred random samples from the DMB halo model posteriors as \textit{blue lines} to illustrate model spread. Note that we normalize the $\langle M|Y_{500c} \rangle$ by $Y_{500c}^{3/5}$ in the top panels to reduce the dynamic range of the y-axis in the plot.  The error bar shows the expected $1\sigma$ uncertainty based on the recent SPT-DESY3 analyses \citep{SPT2}. When constraining the model, we only use bins with $M_{500c}>1.13 \times 10^{14} M_\odot$ reflecting the mass limit of SZ cluster analyses.    In the bottom panels, the \textit{blue symbols} with $1\sigma$ errorbars show the  matter correlation function suppression due to baryonic feedback predicted within this halo model, constrained only by the $\langle M|Y_{500c} \rangle$ measurement in each simulation, with errors dominated by model parameter uncertainties (with priors listed in Table~\ref{tab:Halomodel}).
    The \textit{olive/orange lines} show the matter correlation function suppression measured directly from the two simulations, with errors dominated by cosmic variance. The errorbars show $1\sigma$ uncertainty estimated with $27$ jackknife resamplings.}
    \label{fig:e2e}
\end{figure*}

\section{Summary of main result: validation of the proposed analyses on two different hydrodynamic simulations}

\label{sec:highlevel}
In this paper, we employ the DMB halo model \citep{Schneider19, Giri21, methodpaper} to link measurements of cosmic baryonic physics in the form of mean cluster mass--tSZ scaling relations to the modulation of the total matter distribution.
The relation between tSZ measurements and the effects of feedback on the total matter distribution is complicated by the detection threshold limiting the tSZ measurement to fairly massive clusters ($M_{500c}>1.13 \times 10^{14} M_\odot$\footnote{We verify our result on $M_{500c}>3 \times 10^{14} M_\odot$ corresponding to the mass limit of SPT-Pol \citep{2022ApJ...928...16R} and find consistent result, indicating that the constraints are limited by mass calibration accuracy instead of mass limits.} for the SPT-3G experiment considered here \citep{2022ApJ...928...16R}), which amount to only $7\%$ of the total cosmic matter distribution. Predicting the modulation of the total matter correlation function due to baryonic feedback thus relies on several model assumptions extrapolating the impact of feedback to lower halo masses. However, as we detail in section \ref{scalemass}, galaxy clusters contribute significantly to the baryonic modulation of the cosmic shear signal on the relevant scales, and their impact on the baryonic modulation is expected to be correlated to those of low-mass halos \citep{2023MNRAS.523.2247S}.

To further validate our approach and confirm the appropriateness of extrapolating the model to lower halo masses, we first demonstrate its performance on two different hydrodynamical simulations: \textsc{Magneticum} Pathfinder \citep{2014MNRAS.442.2304H} and \textsc{IllustrisTNG} \citep{2019ComAC...6....2N}. These two simulations differ in many aspects, including different cosmological parameters for initial conditions (WMAP7 \cite{2011ApJS..192...18K} vs Planck15 \cite{2016A&A...594A..13P}), different hydrodynamic treatments (smoothed particle hydrodynamics vs moving-mesh finite volume), different box size ($(352 \hinv \rm{Mpc})^3$, $(205 \hinv \rm{Mpc})^3$), and different mass resolutions ($6.9\times 10^{8}\hinv M_\odot$, $4\times10^{7}\hinv M_\odot$).  While both simulations employ feedback mechanisms linked to \citep{2005Natur.433..604D, 2008ApJ...676...33D}, each has undergone significant development and refinement in subsequent studies \citep{2013MNRAS.436.3031V, 2014MNRAS.442.2304H}. In combination, these differences lead to different predictions in the modulation of matter clustering due to the presence of baryons (see orange and olive lines in figure \ref{fig:hydro}).

This paper considers the mean cluster mass--tSZ relation $\langle M | Y_{500c}\rangle $ as our data vector, where $Y_{500c}$ is the integrated tSZ signal within $R_{500c}$\footnote{We note that the three dimensional $Y_{500c}$ is not a direct observable.  For practical applications, it's crucial to corroborate our findings using the projected  Y measurement. We leave this for future work.}. %
Our data vector is constructed using the 3-D integrated tSZ signal ($Y_{500c}$, see \ref{eq:tsz}) from \textsc{IllustrisTNG}-300 and Box2hr %
of \textsc{Magneticum} Pathfinder \citep{2022MNRAS.517.5303L}. Halos are binned by $Y_{500c}$ to measure their mean masses, mimicking the weak lensing mass calibration approach in data analyses. Halos below the SPT-3G detection threshold are excluded. The remaining bins are assigned $5\%$ uncertainties reflecting the mass calibration uncertainties of the recent joint SPT and DES-Y3 analyses \citep{SPT2}. This data vector constrains our DMB halo model detailed in section \ref{sec:theory} by employing a standard MCMC sampler \citep{2013PASP..125..306F} with a Gaussian likelihood and priors summarized in table \ref{tab:Halomodel}.

We draw one hundred samples from the posterior on the baryon halo model parameters and predict the modulation of the matter correlation function within this model (blue symbols and error bars in figure \ref{fig:hydro}). 
We then compare this with the direct measurement of the matter correlation function suppression from the hydrodynamic simulations (orange and olive lines in figure  \ref{fig:e2e}). We see that galaxy clusters with weak lensing mass calibrations place $\simeq 2\%$ constraints on the modulation of matter clustering at $1\ \hMpc$. Further, the calculation based on galaxy clusters reproduces the directly measured modulation in both hydrodynamic simulations, demonstrating that galaxy clusters contain important information on the modulation of matter clustering due to the presence of baryons.

\section{Theory framework}
\label{sec:theory}
\subsection{A DMB halo model}
\label{sec:1h}
This section shows our implementation of the DMB halo model. 
We note that the distribution profiles of matter components used in this model follow \citep{Schneider19,Giri21}  with additional freedom to account for their variations observed in hydrosimulations. 
For further details and validations, we refer the reader to \citep{methodpaper}. Here, we provide a brief summary.
We model the matter distribution $\rho(r,M_{200c},z, c_{200c})$ around an isolated halo at a given mass ($M_{200c}$) and redshift ($z$)  with three components, reading as 
\begin{eqnarray}
    \rho(r,M_{200c},z, c_{200c}) = \rho_{\rm{gas}} (r,M_{200c},z)+\rho_{\rm{cg}} (r,M_{200c})+\rho_{\rm{clm}}(r,M_{200c},z, c_{200c}), 
\end{eqnarray}
where  $\rho_{\rm{gas}} (r,M_{200c},z)$ describes intracluster medium, $\rho_{\rm{cg}} (r,M_{200c})$ describes stars in the central galaxy of the halo, $\rho_{\rm{clm}} (r,M_{200c},z, c_{200c})$ consists of stars in satellite galaxies and dark matter, and $r$ is the distance to the halo center. 

These components depend on the total stellar fraction, $f_{\rm{star}}$, and the stellar fraction in the central galaxy, $f_{\rm{cg}}$, whose dependence on halo mass is modeled as 
\begin{eqnarray}
\label{eq:star}
f_{\rm{star}} (M_{200c}) = 0.055 \left(\frac{2.5\times10^{11} h^{-1}M_\odot}{M_{200c}}\right)^{\eta_{\rm{star}}}, \\
f_{ \rm{cg}} (M_{200c}) = 0.055 \left(\frac{2.5\times10^{11} h^{-1}M_\odot}{M_{200c}}\right)^{\eta_{ \rm{cg}}},
\end{eqnarray}
where $\eta_{\rm{star}}=\eta$ and $\eta_{\rm{cg}}=\eta+\delta \eta$ %

Following \citep{Giri21}, the gas component $\rho_{\rm{gas}} (r,M_{200c},z)$ is modeled as a cored double power law, written as 
\begin{equation}
\label{eq:rhogas}
    \rho_{\rm{gas}} (r,M_{200c},z) \propto \frac{\Omegab/\Omegam-f_{\rm{star}}(M_{200c})}{\left(1+10 r/R_{200c}\right)^{\beta(M_{200c},z)} \left(1+ r/\left(R_{200c}\theta_{\rm{ej}}\right)\right)^{\left(\delta-\beta\left(M_{200c},z\right)\right)/\gamma}},
\end{equation}
where $\theta_{\rm{ej}}, \delta$, and $\gamma$ are free parameters describing the gas ejection radius and slope, and $\beta (M_{200c},z)$ is modeled as 
\begin{equation}
\label{eq:betam}
    \beta (M_{200c},z) = \frac{3}{1+(M_{c0}(1+z)^{\nu_{\rm z}}/M_{200c})^\mu}, 
\end{equation}
 where $M_{c0}$, $\nu_{\rm z}$, and $\mu$ are free parameters.  %
 The central galaxy component is modeled as a truncated exponential profile,
\begin{equation}
    \rho_{\rm{cg}} (r,M_{200c})=\frac{f_{\rm{cg}}(M_{200c})}{0.06\pi^{3/2}R_{200c}r^2} \rm{exp}\left(-\left(\frac{r}{0.03R_{200c}}\right)^2\right).
\end{equation}

Based on the gravity-only simulation, the collisionless matter component $\rho_{\rm{clm}}$ is expected to follow a truncated NFW profile \citep{nfw}. However, the presence of gas and central galaxies changes the gravitational potential, thus changing the distribution of collisionless matter. We assume that the timescale of this potential change is much longer than that of the dark matter orbiting time scale, thus modeling the effect as an adiabatic process. With these assumptions, the  $\rho_{\rm{clm}}$ can be written as
\begin{eqnarray}
    \rho_{\rm{clm}}(r,M_{200c},z, c_{200c}) &=& \left(1-\Omegab/\Omegam+f_{\rm{star}}(M_{200c})-f_{\rm{cg}}(M_{200c})\right) \rho_{\rm{nfw}}\left(r/\zeta(r,M_{200c},z, c_{200c}), c_{200c}\right)\nonumber\\
    \rho_{\rm{nfw}}(r, c_{200c})&\propto& 
    \frac{1}{\left(\frac{rc_{200c}}{R_{200c}}\right)\left(1+\frac{rc_{200c}}{R_{200c}}\right)^2} \frac{1}{\left(1+\left(\frac{r}{4R_{200c}}\right)^2\right)^2},
\end{eqnarray}
where $R_{200c}$ is the halo radius and $c_{200c}$ is the halo concentration. $\zeta (r,M_{200c},z, c_{200c})$ is the root of
\begin{equation}
    \zeta -1 = 0.3\left(\left(\frac{\int_0^r\rho_{\rm{nfw}}(t,c_{200c}) t^2 dt}{\int_0^r \rho_{\rm{clm}}(t, M_{200c}, z, c_{200c}) t^2 dt + \int_0^{\zeta r} \left(\rho_{\rm{gas}}(t,M_{200c},z)+\rho_{\rm{cg}}(t,M_{200c}) \right)t^2 dt}\right)^2-1\right).
\end{equation}
Finally, $\rho_{\rm{clm}}(r,M_{200c},z, c_{200c})$ and $\rho_{\rm{gas}}(r,M_{200c},z)$ are normalized by the same factor such that 
\begin{eqnarray}
    \int_0^\infty \left(\rho_{\rm{gas}}(r)+\rho_{\rm{cg}}(r)+\rho_{\rm{clm}}(r)\right) 4\pi r^2 dr = \int_0^\infty \rho_{\rm{nfw}}(r) 4\pi r^2 dr.
\end{eqnarray}
\begin{table*}
\begin{tabular}{l l l l}
\hline
\hline
Parameter  & Prior  & Description & Relevant equations \\
\hline
\multicolumn{3}{c}{\textbf{Gas component}} \\ 

$\log_{10} M_{c0}$ & flat(13, 15) & \multicolumn{1}{l}{\multirow{2}{9cm}{The pivot mass scales when the slope of gas profile becomes shallower than $3$.}} & Equation \ref{eq:betam}\\\\

$\nu_{\rm z}$ & flat(-1,1) & Redshift dependence of the pivot mass scale. & Equation \ref{eq:betam}\\

$\mu$ & flat(0,2) & Mass dependence of profile slope.& Equation \ref{eq:betam}\\

$\theta_{\rm{ej}}$ & flat(2,6) & Maximum radius of gas ejection relative to $R_{200c}$. & Equation \ref{eq:rhogas}\\

$\gamma$ & flat(1,4) & Slope of gas profile. & Equation \ref{eq:rhogas}\\

$\delta$ & flat(3,11) & Slope of gas profile. & Equation \ref{eq:rhogas}\\

$\alpha_{\rm{nt}}$ & flat(0.01, 0.4) & Fraction of gas pressure caused by non-thermal processes. & Equation  \ref{eq:thermal}\\ 
\hline
\multicolumn{3}{c}{\textbf{Stellar component}} \\
$\eta$ & flat(0.21,0.36) & Mass dependence of stellar fraction. & Equation.  \ref{eq:star}\\ 

$\delta \eta$ & flat(0.05,0.4) & \multicolumn{1}{l}{\multirow{2}{9cm}{Difference between mass dependence of stellar fraction in central galaxies and total stellar fraction.}} & Equation  \ref{eq:star}\\\\
\hline
\multicolumn{3}{c}{\textbf{Y-M relation}} \\
$\sigma_{\rm{ln}Y}$ & flat(0.1, 0.3) & Scatter of the Y-M relation & log-normal
\\
\hline 
\hline 
\end{tabular}
\centering
\caption{\label{tab:Halomodel} Parameters and priors of our halo model described in section \ref{sec:1h}.} %
\end{table*}

\subsection{Thermal tSZ signal}
The tSZ model implementation follows closely \citep{Baryonpasting, 2014ApJ...792...25N}. %
Assuming hydrostatic equilibrium, the total pressure profile, $P_{\rm{tot}}(r)$, is computed from the gas density profile via 
\begin{eqnarray}
    P_{\rm{tot}}(r) = \int_r^\infty \frac{GM_{\rm{tot}} (<t)}{t^2}\rho_{\rm{gas}}(t, M_{200c},z) dt, \\
    M_{\rm{tot}}(<r) = 4\pi \int_0^r t^2\rho(t, M_{200c},z) dt.
\end{eqnarray} 
This total pressure includes thermal pressure $P_{\rm{thermal}}$ as well as contributions from non-thermal processes, which do not contribute to the tSZ signal.
Following \cite{2014ApJ...792...25N}, we use a simple fitting formula to determine the ratio of thermal pressure and total pressure profile:
\begin{eqnarray}
\label{eq:thermal}
    \frac{P_{\rm{thermal}}}{P_{\rm{tot}}}&=& {\rm{ max}} [0, 1-\alpha_{\rm{nt}} f(z) (r/R_{500c})^{0.8}]\nonumber, \\
    f(z) &=& {\rm{min}} [(1+z)^{0.5}, (6^{-0.8}/\alpha_{\rm{nt}}-1){\rm{tanh}}(0.5z)+1]
\end{eqnarray}
where $\alpha_{\rm{nt}}$ is a free parameter. Finally, we assume that the gas is fully ionized and electrons and protons are in thermal equilibrium within galaxy clusters. This equilibrium assumption is generally valid at the center of galaxy clusters but is questionable at the outskirt \citep{2009ApJ...701L..16R,2015ApJ...808..176A}. Since we focus on the central region of clusters in this paper, we adopt this assumption for the rest of the paper. With the equilibrium assumption, the electron pressure $P_e$ can be computed as 
\begin{eqnarray}
\label{eq:electron}
    P_{e} = \frac{4-2Y}{8-5Y} P_{\rm{thermal}}
\end{eqnarray} 
where $Y=0.24$ is the primordial helium mass fraction. With the electron pressure profile $P_{\rm{e}}$, the thermal tSZ signal $Y_{500c}$ is given by  
\begin{eqnarray}
\label{eq:tsz}
    Y_{500c} = \frac{\sigma_{\rm{T}}}{m_{\rm{e}} c^2} \int_{0}^{R_{500c}} P_{\rm{e}}(r) 4\pi r^2 dr,
\end{eqnarray}
where $\sigma_{\rm{T}}$ is the Thomson cross section, $m_{\rm{e}}$ is the electron mass, and $c$ is the speed of light.
We can then calculate the mean mass of clusters given $Y_{500c}$ bins, assuming a log-normal $Y_{500c}$--M relation. The mean of the relation is given by equation \ref{eq:tsz}
, and the scatter $\sigma_{\rm{ln}Y}$ is a free parameter.

The free parameters of this model and the associated priors assumed in analyses presented in this paper are summarized in table \ref{tab:Halomodel}.

\subsection{Matter Power spectrum and matter correlation function}
\label{sec:MP}
The matter power spectrum $P_{\rm mm} (k,z)$ is computed as a sum of one-halo and two-halo contributions. %
The one halo term $P_{\rm{1H}}(k,z)$  is written as %
\begin{eqnarray}
    P_{\rm{1H}}(k,z) = \int_0^\infty dc_{200c}\int_{M_1}^{M_2} W(M_{200c},k,z,c_{200c})^2 n(M_{200c,z})p(c_{200c}|M_{200c}) dM_{200c} , 
\end{eqnarray}
where we model the halo mass--concentration relation $p(c_{200c}|M_{200c})$ as a log-normal distribution with a scatter $0.11$ dex\footnote{We have checked that increasing the scatter by $50\%$ does not change the result.} %
\citep{2008MNRAS.390L..64D} and a mean following \citep{2012MNRAS.423.3018P}. In the above equation, the lower mass limit $M_1$ is $10^{12} h^{-1} M_\odot$, the upper mass limit $M_2$ is $10^{16} h^{-1} M_\odot$ , the halo mass function $n(M_{200c},z)$ is modeled using the fitting function \citep{Tinker2010}, and $W(M_{200c},k,z, c_{200c})$ is the fourier transform pair of $\rho(r, M_{200c},z, c_{200c})$ which is given by %
\begin{eqnarray}
\label{eq:window}
    W(M_{200c},k,z,c) = \int_0^\infty 4\pi r^2 \frac{\sin(k r)}{kr}\frac{\rho(r, M_{200c}, z, c_{200c})}{\rho_{\rm m}(z)} dr,
\end{eqnarray}
where ${\rho_{\rm m}}(z)$ is the average matter density of the universe.

The two halo term $P_{\rm{2H}}(k,z)$ is given by the product of the linear power spectrum and the square of the mean biases, written as,
\begin{eqnarray}
\label{eq:salmon}
    P_{\rm{2H}}(k,z) &=& P_{\rm{lin}}(k,z) \left(I(M_1,k,z)+ A(M_1,z) \right)^2 \\
    I(M_1,k,z) &=& \int_0^\infty dc_{200c}\int_{M_1}^{\infty} W(M_{200c},k,z, c_{200c})n(M_{200c})p(c_{200c}|M_{200c}) b(M_{200c},z) dM_{200c} \nonumber\\
    A(M_1,z) &=& 1-I(M_1, k=0,z) \nonumber,
\end{eqnarray}
where $P_{\rm{lin}}(k,z)$ is the linear power spectrum of matter fluctuation, the linear halo bias $b(M_{200c},z)$ is modeled with the fitting function of \citep{Tinker2010}, $I(M,k,z)$ is the standard mean bias, and $M_1$ is the lower limit of the integral setting to $10^{12} \hinv M_\odot$. Because the matter distribution is unbiased with respect to itself, the mean biases must approach $1$ on large scales. While the modeling of $b(M,z)$ and $n(M,z)$ in \citep{Tinker2010} guarantees that $I(0,0,z)=1$, $I(M_1,0,z)\neq0$ when $M_1\neq 0$. To mitigate this problem, we follow \citep{Mead2020} to add $A(M_1,z)$ in the mean bias calculation to ensure a proper asymptotic value at $k=0$. %

So far, our procedure of calculating the matter power spectrum adheres to the standard halo model framework. 
Applying this to the DMB halo model provides the necessary flexibility and physical ingredients to describe the baryonic impact of the matter power spectrum \citep{Giri21}, yet it approximates the gravity-induced non-linear matter power spectrum at the $\simeq 20\%$ accuracy compared to accurate gravity-only simulation \citep[e.g.][]{2015MNRAS.454.1958M}. 
Given that the non-linear matter clustering induced by gravity can be accurately modeled with N-body simulations, it is natural to differentiate between non-linear clustering and baryonic impacts. Here, we adopt the approach employed in \citep{2015MNRAS.454.1958M, Mead2020} that uses the halo model to describe the ratio of the matter power spectrum in the presence of baryons to that in a gravity-only
universe, and uses more accurate models to describe the non-linear matter clustering. %
That is, our model of the matter power spectrum $P_{\rm mm}(k,z)$ is 
\begin{eqnarray}
    P_{\rm mm}(k,z) = P_{\rm SIM, G}(k,z) \frac{P_{\rm HM, B}(k,z)}{P_{\rm HM, G}(k,z)},
\end{eqnarray}
where $P_{\rm HM, B}(k,z)$ and $P_{\rm HM, G}(k,z)$  are halo model power spectra obtained from applying the procedure described in section \ref{sec:MP} on the DMB halo model $\rho(r,M_{200c},z, c_{200c})$ and the NFW profile $\rho_{\rm NFW}(r, c_{200c})$ respectively. The ratio of $P_{\rm HM, B}(k,z)$ and $P_{\rm HM, G}(k,z)$ describes the baryonic impact on the matter power spectrum. $P_{\rm SIM, G}(k,z)$ is the accurate matter power spectrum in a gravity-only universe, which can come directly from gravity-only simulation \citep[e.g.][]{Aemulusnu, Euclidemulator2} or accurate fitting functions \citep{Mead2020, halofit2, halofit}. For simplicity, we adopt the Eisenstein and Hu transfer function \citep{EH98} and Halofit implemented in \citep{jaxcosmo} as $P_{\rm mm, G}(k,z)$ for this work. With the matter power spectrum, one can then calculate the matter correlation function $\xi_{\rm mm}(r)$ via Fourier transform, and cosmic shear correlation functions $\xi_{+/-}(\theta)$ via Limber approximation, Hankel transform, and an assumed redshift distribution. 

\begin{figure*}
\centering
    \includegraphics[width =0.95\textwidth]{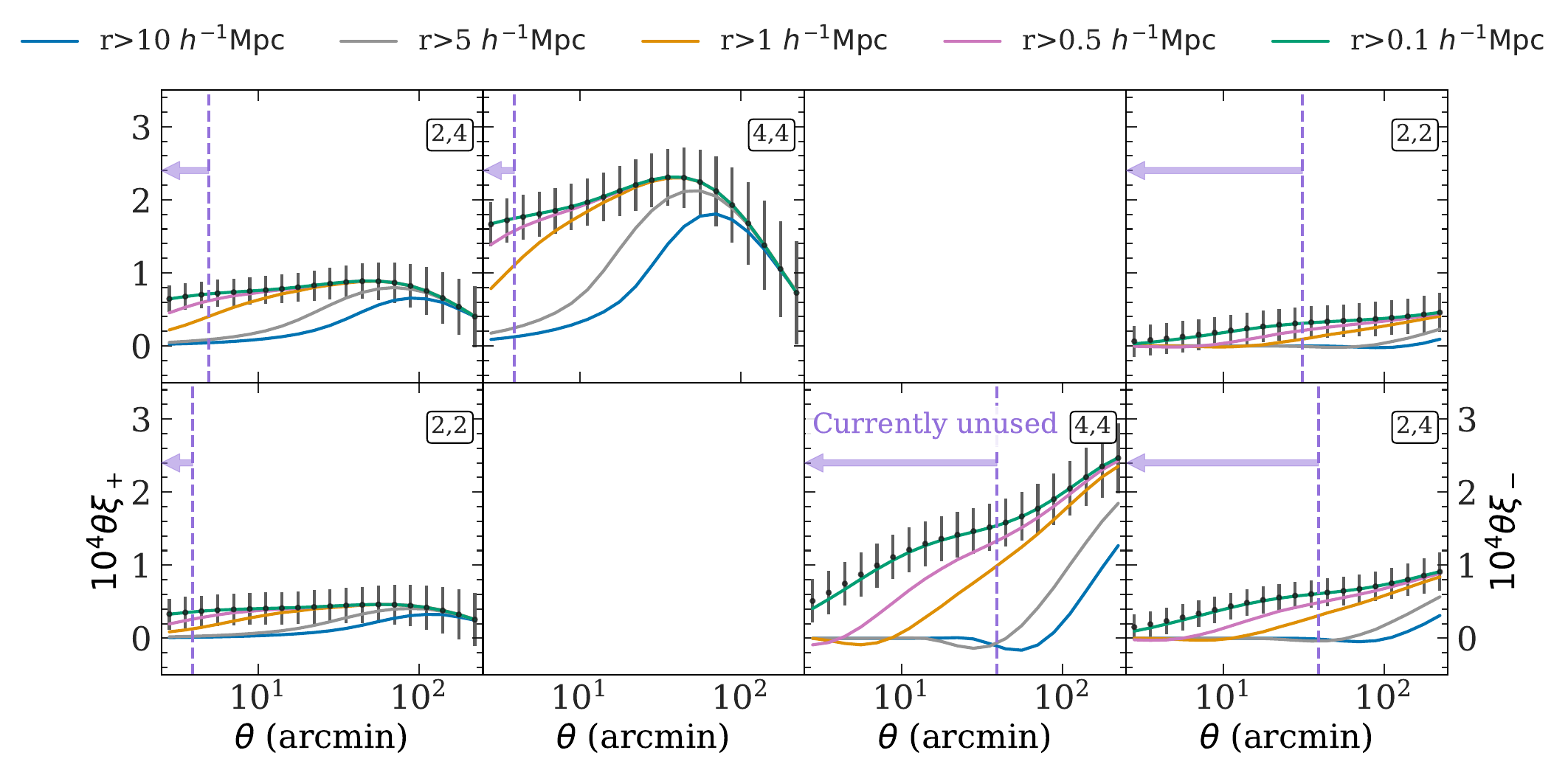}
    \caption{Contributions from different physical scales to the cosmic shear 2-point correlation functions. Model predictions for $\xi_+$ (upper left) and $\xi_-$ (lower right) in the DES-Y3 tomographic bins are shown as \textit{black symbols}, with errorbars corresponding shape noise and cosmic variance of DES-Y3. The \textit{blue/grey/orange/pink/green line} are obtained by setting $\xi_{\mathrm{mm}}(r) = 0$ for $r < 10/5/1/0.5/0.1\, \hinv \rm{Mpc}$ in the computation of $\xi_\pm$. The angular scales smaller than the purple dashed lines are excluded in the DES-Y3 $\Lambda$CDM-optimized analyses with more aggressive scale cuts than the 3x2pt analyses \citep{Y3_Amon, Y3_Secco}. This plot shows the auto and cross-correlations of two DES-Y3 redshift bins corresponding to a mean redshift of $0.5$ and $0.93$, respectively.}
    \label{fig:rcontrib}
\end{figure*}
\begin{figure*}
\centering
    \includegraphics[width =0.8\textwidth]{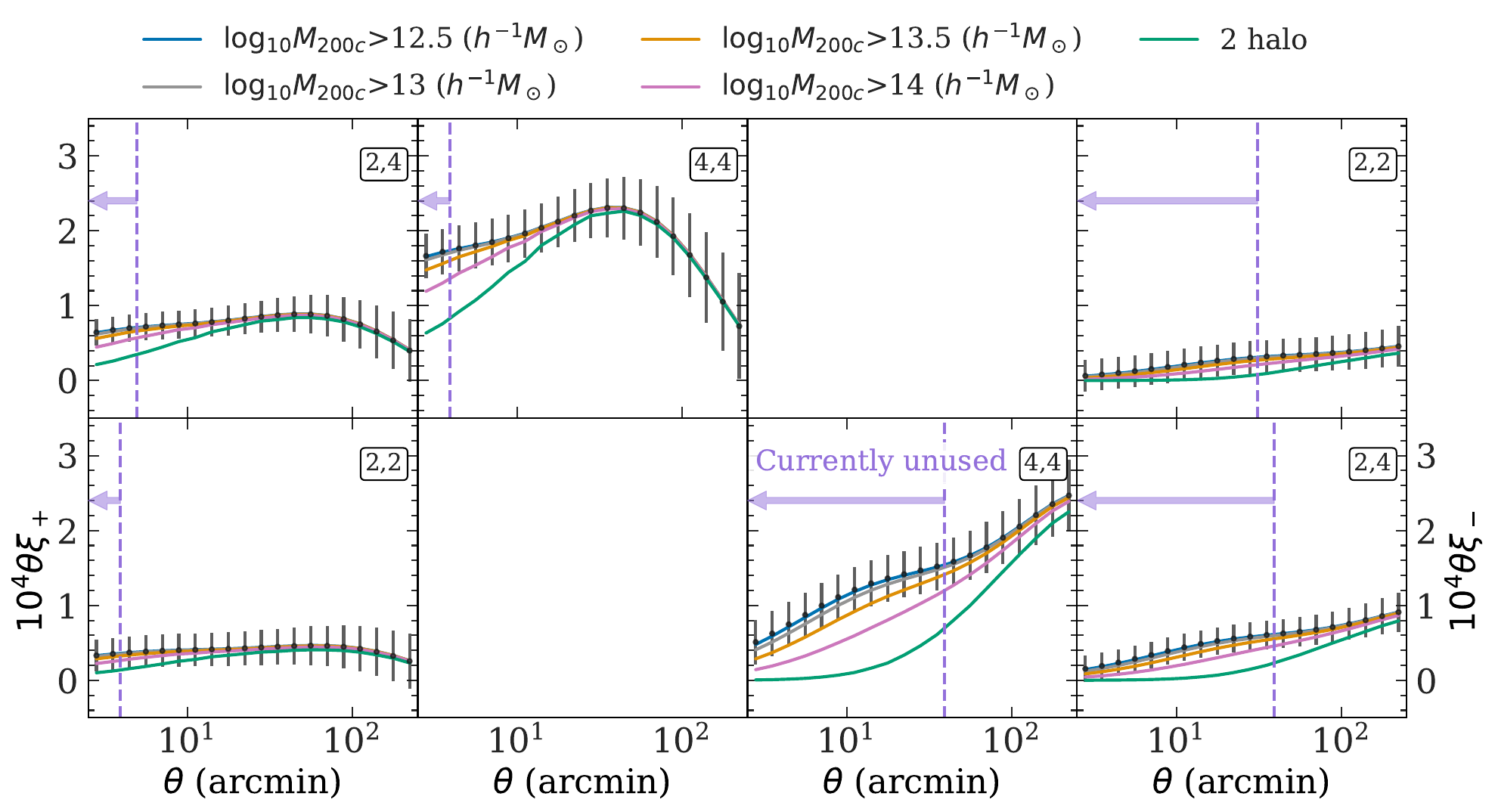}
    \caption{Contributions from halos with different masses to the cosmic shear 2-point correlation functions. Model predictions for $\xi_+$ (upper left) and $\xi_-$ (lower right) in the DES-Y3 tomographic bins are shown as \textit{black symbols}, with errorbars corresponding shape noise and cosmic variance of DES-Y3. The \textit{blue/grey/orange/pink/green line} are obtained by excluding one-halo contribution in $\xi_{\mathrm{mm}}$ from halo with mass $M_{200c}< 10^{12.5}/10^{13}/10^{13.5}/10^{14}/\infty \, \hinv M_\odot$ in the computation of $\xi_\pm$. The angular scales smaller than the purple dashed lines are excluded in the DES-Y3 $\Lambda$CDM-optimized analyses with more aggressive scale cuts \citep{Y3_Amon, Y3_Secco}. This plot shows the auto and cross-correlations of two DES-Y3 redshift bins corresponding to a mean redshift of $0.5$ and $0.93$, respectively.}
    \label{fig:Mscontrib}
\end{figure*}

\section{Disecting the cosmic shear signal}
\label{scalemass}
\begin{figure*}
\centering
    \includegraphics[width =0.8\textwidth]{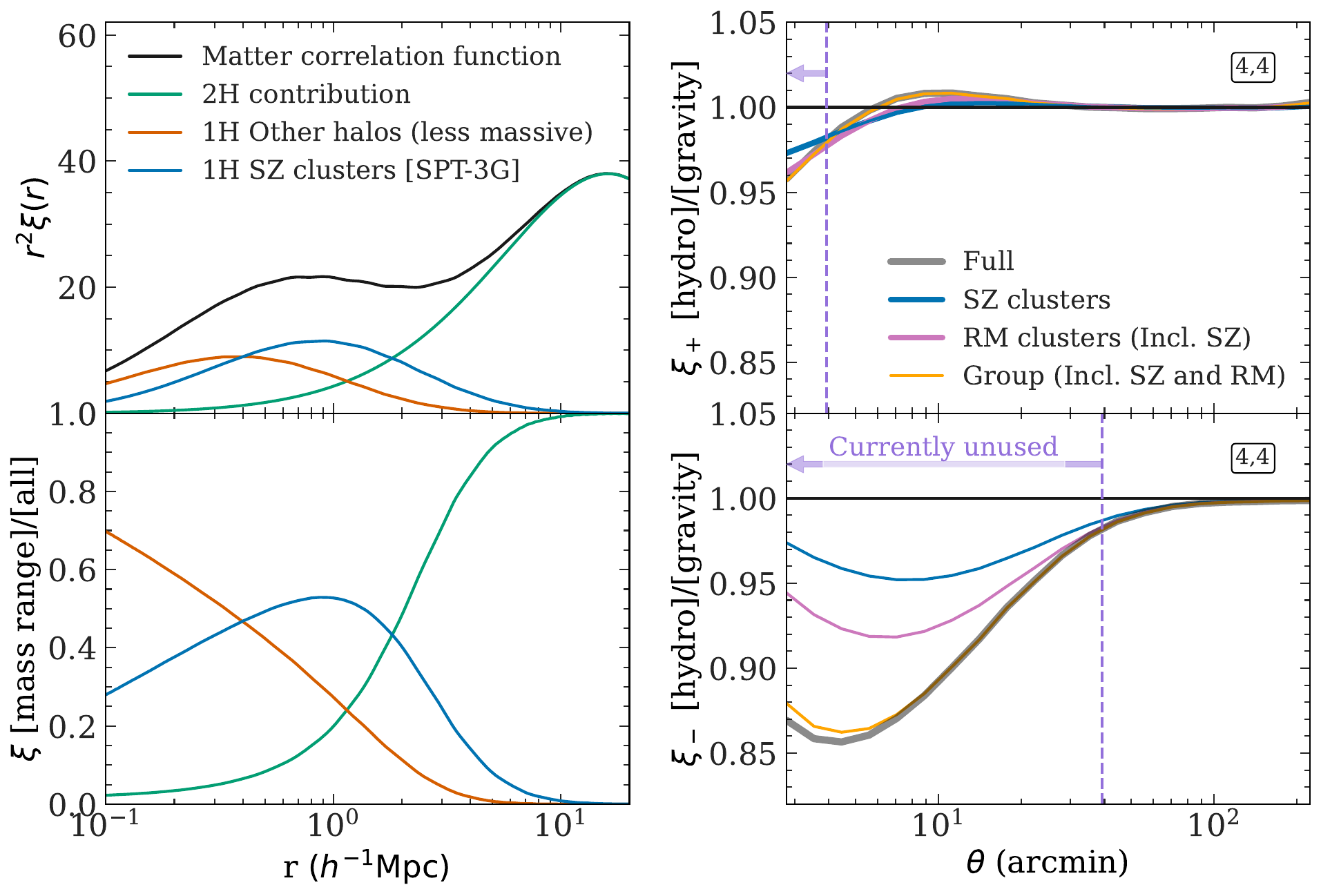}
    \caption{Contribution of different mass ranges to the matter correlation function (left) and their contribution to the baryonic suppression of shear 2-point statistics (right). The first column shows the matter correlation function, and the bottom left panel shows the fractional contribution of a specific mass range for the one-halo term contribution from SZ-selected clusters (\textit{blue line}), the one-halo term contribution from halos with mass below the SZ detection threshold (\textit{orange line}), and the complete two-halo term (\textit{green line}). The second column (black line) shows the baryonic feedback suppression of $\xi_+$ (top) and  $\xi_-$ (bottom) at the highest redshift bin of DES-Y3. The \textit{blue}, \textit{pink}, and \textit{yellow} lines show the same suppression but assuming baryonic feedback affects only SZ-selected clusters,  \redmapper-selected clusters, and group-scale halos, respectively. Note that here, we assume different cluster samples are complete above a certain mass threshold. That is, SZ-selected clusters are a subset of  \redmapper-selected clusters, which are a subset of group-scale halos.
    }
    \label{fig:Mcontrib}
\end{figure*}
With the DMB halo model, we study the contribution of different physical scales and halo masses to cosmic shear correlation functions $\xi_{+/-}(\theta)$ assuming a DES-Y3-like survey \citep{DESY3}. In this section, we adopt a Planck cosmology \citep{2020A&A...641A...6P} and halo model parameters following contraints from \citep{Grandis2023}: $\log_{10} M_{c0}=14.38$, $\nu_{\rm z}=0$, $\mu=0.6$ , $\theta_{\rm{ej}}=3.0$, $\gamma=1.2$, $\delta=6.0$, $\alpha_{\rm{nt}}=0.18$, $\eta=0.24$, $\delta \eta=0.08$.

In figure \ref{fig:rcontrib}, we compute the cosmic shear correlation functions assuming a DES-Y3 redshift distribution. We further break the correlation functions into contributions from different comoving scales in configuration spaces. These scale cuts are implemented by Fourier transforming the matter power spectrum into the 3D matter correlation function, applying a top-hat filter, and Fourier transforming back to the matter power spectrum. The Fourier transform is done with FFTLog algorithm implemented in mcfit\footnote{ \url{https://github.com/eelregit/mcfit}.}.
The plot shows that the cosmic shear correlation function is sensitive to scales deep in the one-halo regime.  Specifically, one has to include scales down to $0.5\ h^{-1}\rm{Mpc}$ to achieve modeling accuracy better than $1\sigma$ at scales within DES-Y3 scale cuts in the two shown redshift bins, corresponding to a mean galaxy redshift of $0.5$ and $0.93$ respectively. Further, at $2.5$ arcmin scales and the considered redshift ranges, we find that cosmic shear is insensitive to the matter correlation function at scales less than $0.1\ \hinv \rm{Mpc}$, making it robust to the modeling of stellar mass in and around central galaxies. Finally, the contribution from different halo mass ranges in cosmic shear correlation functions is shown in figure \ref{fig:Mscontrib}. Interestingly, we see that including halos with mass ($M_{200c}$) greater than $10^{14}\ \hinv M_\odot$ already leads to model deviations less than $1\sigma$ at scales within DES-Y3 scale cuts, indicating the significant contribution of matter in cluster environments to total cosmic shear signals. This finding is consistent with those in \citep{2007NJPh....9..446T}, where the authors performed a similar calculation in Fourier space.

So far, we have shown that galaxy clusters contribute significantly to the cosmic shear correlation functions. To illustrate the potential of SZ galaxy clusters for constraining baryonic impacts on cosmic shear, we separate the matter correlation function into three components: one-halo contribution from galaxy clusters assuming SPT-3G mass limit ($M_{500c}>1.13 \times 10^{14} M_\odot$ \citep{2022ApJ...928...16R}), one-halo contribution from halos below the mass limit, and two-halo contribution, with relative contributions to the total matter correlation function shown in the left column of figure \ref{fig:Mcontrib}. Consistent with the discussion of figure~\ref{fig:Mcontrib}, SZ clusters dominate the matter clustering signal on scales relevant to cosmic shears. %
In the right column of figure \ref{fig:Mcontrib} we quantify the contribution of baryonic feedback from the SZ cluster mass range to the total baryonic modulation of cosmic shear. Specifically, the blue line in the right column of figure \ref{fig:Mcontrib} shows the modulation of cosmic shears, assuming that the baryonic effect only affects SZ clusters\footnote{For simplicity, we assume the mass limit of SZ clusters to be redshift-independent and focus on the shear correlation function at the fourth redshift bin of DES-Y3.}. 
Comparing this to the total cosmic shear modulations (black line) shows that SZ clusters capture almost all baryonic effects in $\xi_+ $ and about one-third in $\xi_-$ down to $2.5'$ scales. This indicates that constraining the modulation of matter clustering below DES-Y3 scale cuts with SZ clusters requires extrapolations of the baryon suppression model to lower halo masses, despite the significant contribution of SZ clusters to the cosmic shear signal.
While the accuracy of this extrapolation has has been validated in two hydrodynamic simulations figure \ref{fig:e2e}, we also explore the impact of extending the DMB halo model calibration to lower halo masses in the right column of figure \ref{fig:Mcontrib}. We show the contribution of baryonic feedback from different halo mass ranges to the total baryonic modulation of cosmic shear, specifically examining \redmapper{} clusters from the DES \citep{6x2t+N} (magenta line) and the Tinker groups from SDSS \citep{Tinkergroup} (yellow line). For simplicity, we assume a single mass cut of $M_{200c}>5\times10^{13} \hinv M_\odot$ for \redmapper{} clusters and $M_{200c}>5\times10^{12} \hinv M_\odot$ for Tinker groups. Extending the selection threshold to lower masses removes the need for extrapolation, but we caution that modeling the selection function of these samples can be challenging. We also note that all samples fit DMB model parameters with the same parameterization of the mass dependence, albeit over different mass ranges, and are thus not immune to the mass scaling assumptions in Eqs. \ref{eq:rhogas} and \ref{eq:betam}.

\begin{figure*}
\centering
    \includegraphics[width =0.95\textwidth]{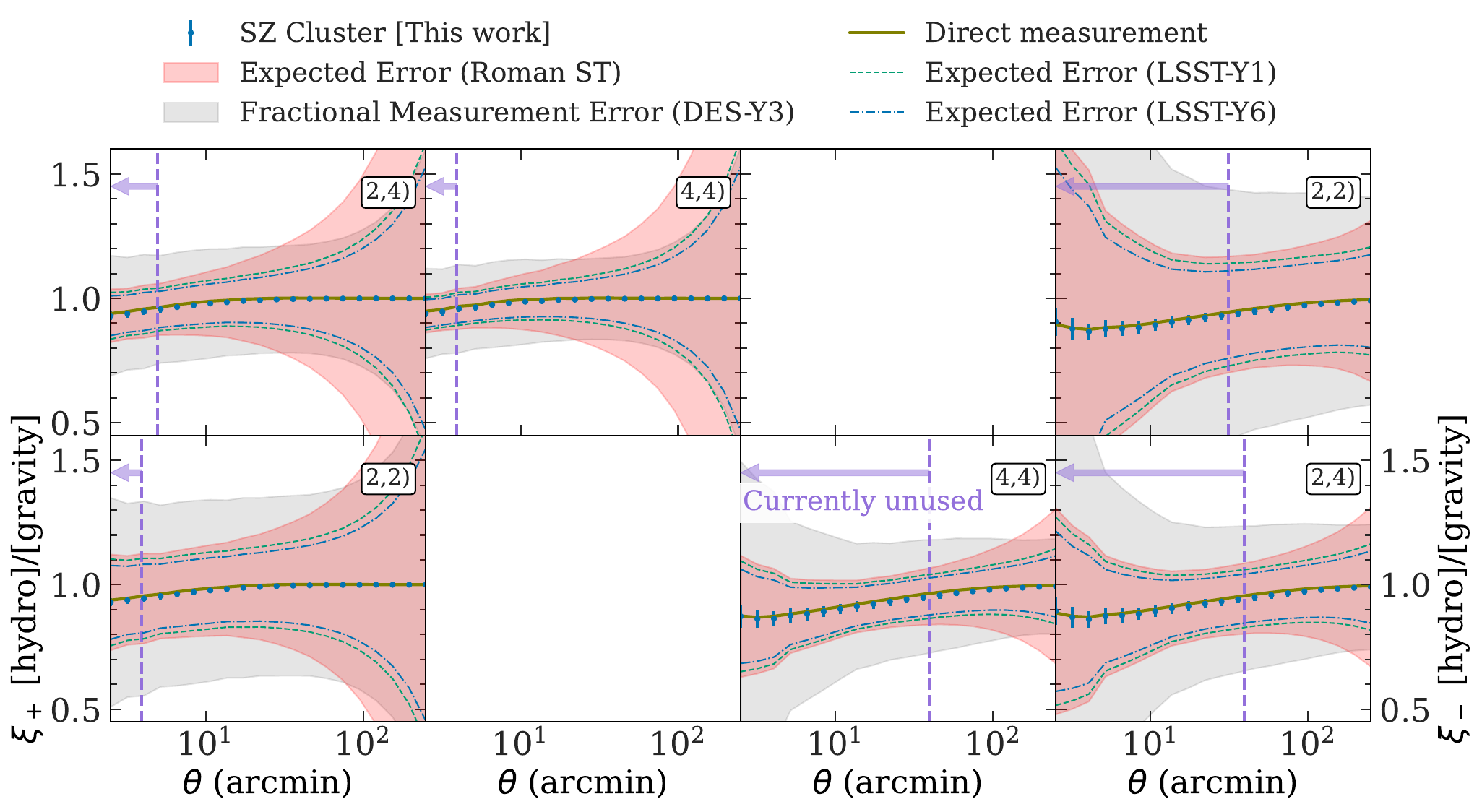}
    \caption{Predictions of cosmic shear modulations due to baryonic feedback based on figure \ref{fig:e2e}. Predictions based on direct measurements in the Magneticum simulation for $\xi_+$ (upper left) and $\xi_-$ (lower right) in the DES-Y3 tomographic bins are
shown as olive lines. Dots with $1\sigma$ error bars show the DMB halo model prediction based on SZ cluster measurements in the simulations. The gray bands show the $1\sigma$ fractional error of the DES-Y3 measurement \citep{DESY3}, the red bands show that of the high latitude survey of the Roman Space Telescope \citep{RomanTim}, the green dashed lines show that of LSST-Y1 \citep{2022MNRAS.509.5721F}, and the blue dashed lines show that of LSST-Y6 \citep{2024MNRAS.527.9581F}. For simplicity, we assume no redshift evolution of baryonic impact on matter correlation functions.}
    \label{fig:e2e_shear}
\end{figure*}

\begin{figure}
\parbox{.45\linewidth}{%
\includegraphics[width=\linewidth]{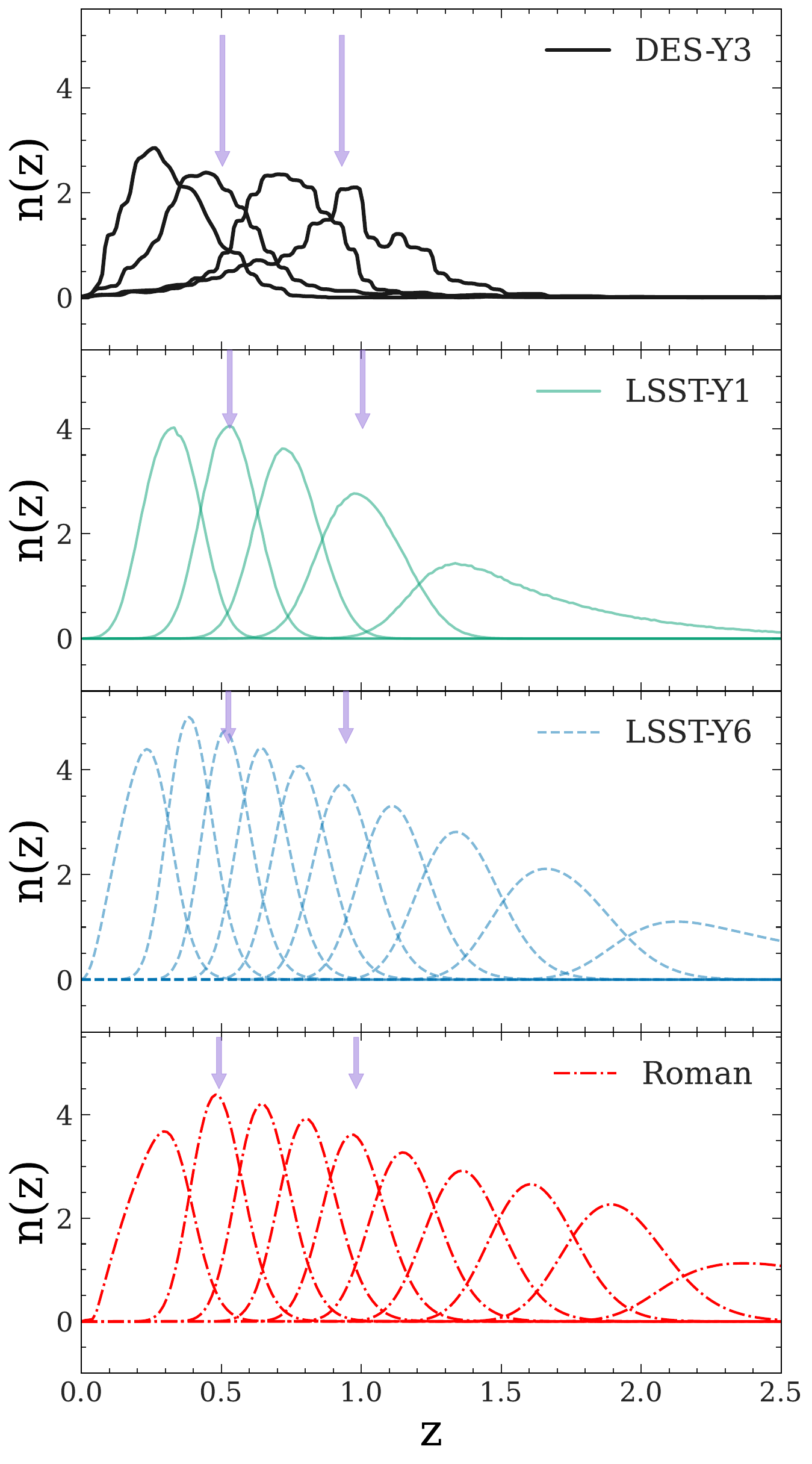}
}\hfill
\parbox{.7\linewidth}{\includegraphics[width=\linewidth]{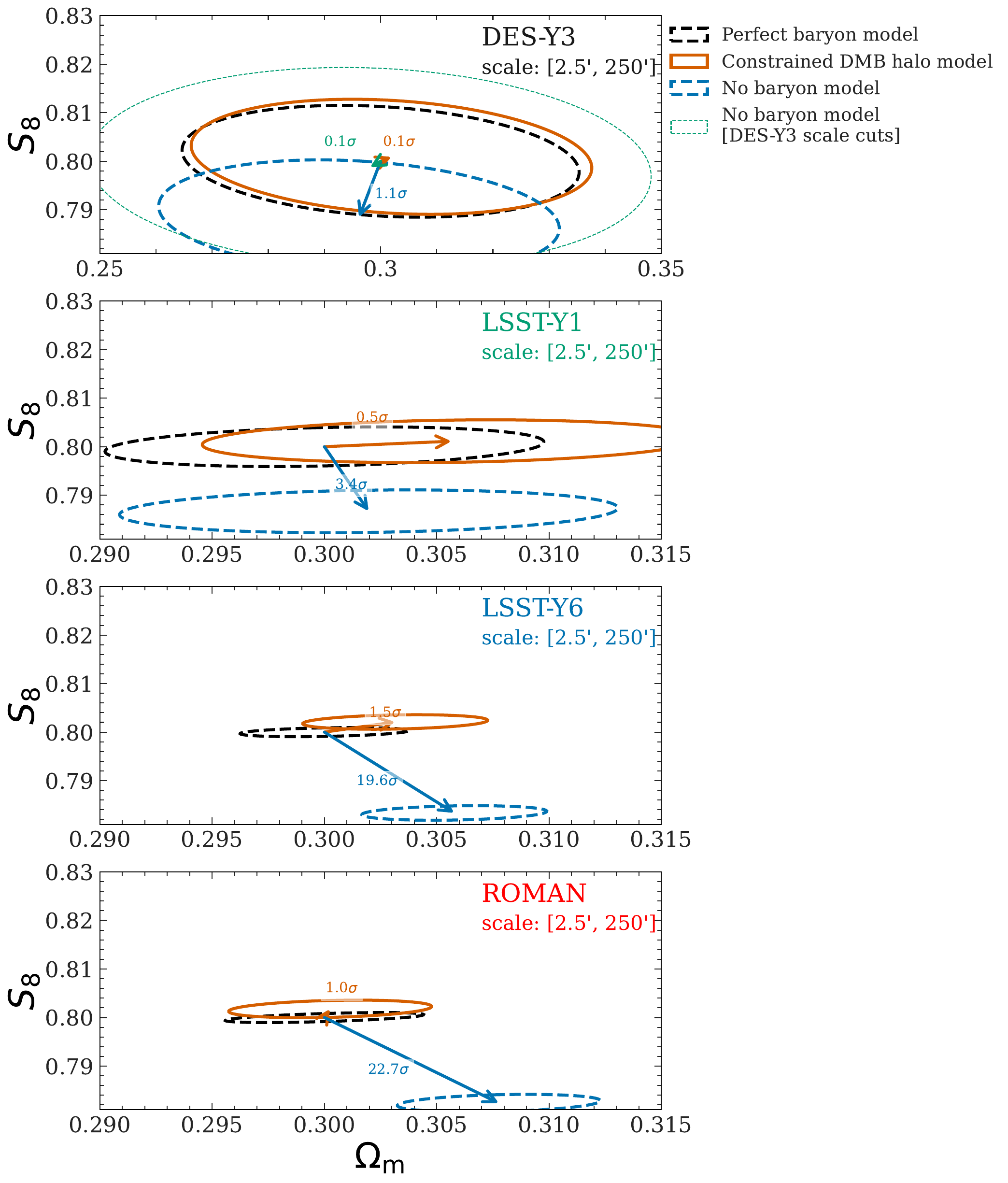}}
\par
\parbox[t]{.475\linewidth}{\caption{Simulated redshift distribution of Roman (red), LSST-Y1 (green) and LSST-Y6 (blue) source galaxies. In comparison, the redshift distribution of DES-Y3 is shown as black lines. Bins indicated by arrows are used to produce figure \ref{fig:e2e_shear}.\label{fig:zdist}}}\hfill
\parbox[t]{.475\linewidth}{\caption{Forecasted constraints on $S_8$ and $\Omegam$ of cosmic shears from $2.5'$ to $250'$ measured by DES-Y3 (top), LSST-Y1 (the second panel), LSST-Y6 (the third panel), and Roman (the fourth panel). Contours show the $1\sigma$ region. The black dashed lines show the constraints from a hypothetical perfect model of the baryonic feedback. The orange contours correspond to the DMB halo model constrained by tSZ clusters. The blue dashed lines show the constraints from the model only considering gravity. As a comparison, the green dashed line in the first panel shows the constraints with DES-Y3 fiducial scale cuts \citep{DESY3}. Systematic biases are indicated by arrows with the value quantified relative to the uncertainty of the biased analysis, e.g., orange arrow relative to orange contour.}
  \label{fig:Fisher}}
\end{figure}

\section{Outlook}
\label{sec:Outlook}
\begin{table}
	\centering
	\vspace{1cm}

\resizebox{\columnwidth}{!}{\begin{tabular}{cccccc}
\hline 
Parameter & Fiducial value & Prior (DES-Y3)& Prior (LSST-Y1)& Prior (LSST-Y6)& Prior (Roman)\tabularnewline
\hline 
\hline
\multicolumn{6}{c}{\bf{Cosmological Parameters}}  \tabularnewline
$\Omega_{\rm m}$ & 0.31& \multicolumn{4}{c}{$\mathrm{U}[-\infty,  \infty]$} \tabularnewline
$S_{\rm 8}$ & 0.8 &\multicolumn{4}{c}{$\mathrm{U}[-\infty,  \infty]$}   \tabularnewline
$\Omega_{\rm b}$  & $0.049$& \multicolumn{4}{c}{$\mathrm{G}[0.049,  0.01]$}\tabularnewline
$n_{\rm s}$ &$0.95$& \multicolumn{4}{c}{$\mathrm{G}[0.95,0.012]$}   \tabularnewline
$h$  & $0.672$& \multicolumn{4}{c}{$\mathrm{G}[0.672,  0.1]$}\tabularnewline
\hline 
\multicolumn{6}{c}{\bf{Intrinsic Alignment}}  \tabularnewline
$A_{\rm{IA}}$  & $0.1$& \multicolumn{4}{c}{$\mathrm{G}[0.1,3]$}\tabularnewline
$\eta_{\rm{IA}}$  & $0.0$& \multicolumn{4}{c}{$\mathrm{G}[0.0,3.8]$}\tabularnewline
\hline 
\multicolumn{6}{c}{\bf{Shear Calibration}}  \tabularnewline
$m^i$  & $0.0$& $\mathrm{G}[0,0.008]$&$\mathrm{G}[0,0.013]$&$\mathrm{G}[0,0.003]$&$\mathrm{G}[0,0.002]$ \tabularnewline
\hline 
\multicolumn{6}{c}{\bf{Photo-$z$ Bias}}  \tabularnewline
$\Delta_z^i$  & $0.0$& $\mathrm{G}[0,0.011]$&$\mathrm{G}[0,0.002]$&$\mathrm{G}[0,0.001]$&$\mathrm{G}[0,0.002]$ \tabularnewline
\hline
\hline 
\end{tabular}}
    \caption{The parameters varied, their fiducial values, and their priors in the fisher analysis detailed in section \ref{sec:Outlook}. Priors are either uniform (U) or normally-distributed, $\mathrm{G}(\mu,\sigma)$, where $\mu$ is the mean and $\sigma$ is the scatter. $n_s$ prior is chosen to be three times wider than the posterior of \citep{2020A&A...641A...6P}. $h$ and $\Omegab$ priors are chosen to mimic the informative flat priors on these parameters adopted by DES \citep{DESY3}.} \label{table: priors}
\end{table}

\begin{table*}[]
\centering
\begin{tabular}{llllll}
\hline
\hline
                                    &DES-Y3 &        LSST-Y1 & LSST-Y6 & Roman &\\ \hline
$S_8$ systematic bias           & $0.08\sigma$ &  $0.26\sigma$       & $1.40\sigma$        & $0.98\sigma$    &   \\ \hline
$S_8$ increased uncertainties & $0.03\sigma$&   $0.08\sigma$ ($0.07\sigma$)   &   $0.60\sigma$ ($0.46\sigma$)     & $0.77\sigma$  ($0.64\sigma$)    \\ \hline  
$S_8$ bias (no baryon model) &$1.00\sigma$&$3.26\sigma$&$17.87\sigma$&$17.23\sigma$\\ \hline
\hline
\end{tabular}
\centering
\caption{\label{tab:param} Summary of systematic biases and increased uncertainties on $S_8$ in the cosmic shear analyses using the constrained DMB halo model relative to the statistical uncertainties marginalized over the parameters in table~\ref{table: priors}. Numbers in the parenthesis are based on expected constraints of SPT-3G clusters with mass calibrated from next generational weak lensing measurements \citep{vogt}. In comparison, the last row shows the $S_8$ biases in the scenario when the real universe is like the Magneticum simulation and baryonic feedback is completely ignored in the analysis. This forecast uses the full range of angular scales from 2.5-250 arcmin.
} %
\end{table*}

In this paper, we have demonstrated that employing the DMB halo model and the $\langle M | Y_{500c}\rangle $ measurement of SZ clusters can precisely and accurately constrain the matter clustering modulation (figure \ref{fig:e2e}). We explore the implications of this precision and accuracy for both current and future surveys. Assuming the DES-Y3 redshift distribution, figure \ref{fig:e2e_shear} shows the predicted modulation of cosmic shears by Magneticum simulations (olive lines)\footnote{We assume no redshift evolution in the matter clustering modulation for simplicity.}. Predictions using the DMB halo model, constrained by $\langle M | Y_{500c}\rangle$ with the mass calibration accuracy from the SPT-DESY3 joint analysis \citep{SPT2}, are represented by blue dots with error bars showing $1\sigma$ uncertainties. We show the statistical uncertainties (error bars of blue dots, \textit{DMB  uncertainties} ($\rm{cov}_{\rm DMB}$) hereafter) and systematic biases (differences between blue dots and olive lines, \textit{DMB  misspecification} ($\Delta d_{\rm DMB}$) hereafter) of the DMB halo model constrained by SZ clusters to the measurement errors ($\rm{cov}_{\rm stat}$\footnote{While the plot only shows the diagonal terms, we include non-diagonal elements of $\rm{cov}_{\rm DMB}$ estimated from the DMB MCMC samples in the calculation below.}) of DES-Y3 (grey-shaded band) and the expected measurement errors of LSST-Y1, LSST-Y6, and Roman High Latitude surveys (green dotted/blue dashed/red-shaded bands). We model the measurement errors of future surveys as a Gaussian covariance matrix generated with a code detailed in \citep{cosmolike2016}, using the source redshift distributions, shape noise, source number density, and survey areas defined in \citep{RomanTim,2022MNRAS.509.5721F,2024MNRAS.527.9581F}. The assumed redshift distributions are shown in figure \ref{fig:zdist}. For simplicity, we assume the same angular binning as that of DES-Y3.  

Since different surveys shown in figure~\ref{fig:e2e_shear} have different redshift binning, it is essential to quantify the performance of the baryon mitigation method proposed here with a metric that combines all redshift bins. We perform Fisher forecasts \citep[see, e.g.][for a review]{Huterer:2023mmv} to quantify the systematic biases and increased uncertainties in $S_8=\sigma_{\rm 8} (\Omega_{\rm m}/0.3)^{0.5}$ due to DMB misspecification and DMB parameter uncertainties. These analyses assume a $\Lambda$CDM cosmological model and marginalize over observational and astrophysical systematics, including shear calibration, photometric redshift uncertainties, and intrinsic alignments. 
The constraining power on $S_8$ of these surveys can degrade when considering a flexible model of cosmic shear modulation. To make the most conservative assessment, we fix the DMB model parameters when performing the Fisher forecasts. Our detailed implementation of these observational and astrophysical systematics is described in \citep{methodpaper}, and the fiducial parameter values and the priors for the analysis are summarized in table \ref{table: priors}. For each survey, we calculate the $1\sigma$ 1D-marginalized constraints on $S_8$. We can then quantify the significance of the DMB uncertainties relative to the measurement errors by calculating the increased $1\sigma$ uncertainties of $S_8$ due to DMB uncertainties. This is achieved by adopting $\rm{cov}_{\rm stat}+\rm{cov}_{\rm DMB}$ as the covariance matrix in the Fisher forecast instead of using $\rm{cov}_{\rm stat}$. 
 To quantify the systematic biases on $S_8$ due to DMB misspecification, we further use the Fisher bias formalism \citep[see, e.g.][for a review]{Huterer:2023mmv} to transform the measured DMB misspecification $\Delta d_{\rm DMB}$ into systematic biases in $S_8$ constraints. The result is summarized in table \ref{tab:param} and figure \ref{fig:Fisher}\footnote{We note that the forecast results depend on prior choices for cosmology and nuisance parameters. In this analysis, we adopt a Gaussian prior on $\Omegab$, $h$ to mimic the informative flat priors on these parameters adopted by DES. The similarity of LSST-Y6 and Roman constraining power on $S_8$ and $\Omegam$ is driven by differences in shear calibration priors and differences in posteriors of other cosmological parameters.}. We find that combining SZ clusters and the DMB halo model can constrain cosmic shear modulation to an accuracy that leads to less than $0.2\sigma$ biases in $S_8$ for DES-Y3 and LSST-Y1. This bias exceeds $1\sigma$ for Roman and LSST-Y6, indicating a need for further investigations on this systematic bias's origin and mitigation method. This systematic bias can come directly from the DMB halo model's limited accuracy or the inaccurate DMB halo model parameters constrained from high mass halos. The latter will be further reduced when performing a joint analysis of cosmic shears and SZ clusters, where cosmic shears provide significant signal-to-noise to constrain DMB halo model parameters.  Hence, detailed simulated analyses will be required to check whether the systematic bias on $S_8$ can be improved in a joint analysis. We leave this for future work. 
 
 Regarding the statistical uncertainties, we find that SZ clusters can constrain cosmic shear modulation with a precision that leads to less than $0.1\sigma$ increases in $S_8$ uncertainties in DES-Y3 and LSST-Y1. However, this level of precision can lead to more than $0.5\sigma$ increases in $S_8$ uncertainties for LSST-Y10 and Roman. To better understand whether this uncertainty may be reduced with future weak lensing mass calibrations of SZ clusters, we repeat the analysis assuming a mass uncertainty of $2\%$ according to \citep{vogt}, which forecasts an SPT-3G cluster sample and a Euclid-like weak lensing survey with a source galaxy number density of $30\ \rm{arcmin}^{-2}$. The result is shown in the parenthesis in table \ref{tab:param} and figure \ref{fig:Fisher}. The increased precision of mass calibration further reduces the $S_8$  uncertainties, but the contribution of the DMB halo model uncertainties to the total error budget in LSST-Y10 and Roman is still significant. We note that this statistical uncertainty can be reduced with satellite mass constraints in SZ clusters \citep[similar to][for X-ray clusters]{Grandis2023} and includes clusters with lower detection significance.  We leave these investigations for future work. 

So far, our tests assume a fixed set of cosmological parameters. In practice, weak lensing mass calibrations of galaxy clusters also depend on cosmology and share several systematics with cosmic shears, such as photometric redshift biases and multiplicative biases of shape measurements. Therefore, a combined analysis of cluster lensing and cosmic shear to constrain cosmological parameters while marginalizing the baryon parameters in the DMB halo model is the natural framework for obtaining rigorous constraints. It not only helps in tightening constraints on cosmological parameters but also allows for a consistent model of different data. While we do not explicitly test this combined analysis in this paper, our result indicates that such analysis will work at the level of accuracy required for DES-Y3 and LSST-Y1 analyses. The SZ cluster constraints with current survey data are expected to be precise and accurate enough that the small-scale cosmic shear constraints on the cosmological parameters will not be significantly degraded by marginalizing baryonic physics uncertainties.

\section{Conclusion}
\label{sec:conc}

Upcoming weak lensing surveys will achieve percent-level precision in measuring the total matter distribution. This distribution is shaped by gravity and astrophysical processes, among which baryonic feedback from galaxy formation plays a crucial role. Accurate and precise constraints of baryonic feedback and its impact on the matter distribution are thus essential to weak lensing cosmological studies. This paper shows that tSZ-selected galaxy clusters with mass calibrated by current weak lensing data (DES-Y3) can constrain cosmic shear modulations due to baryonic feedback with statistical and systematic errors subdominant to DES-Y3 and LSST-Y1 measurement errors (figure  \ref{fig:Fisher} and table \ref{tab:param}). This constraint is achieved with a flexible DMB halo model. 

Our main findings are as follows:
\begin{itemize}
    \item In figure \ref{fig:rcontrib}, we show that cosmic shears are sensitive to the matter correlation function on scales less than $1\ \hMpc$. Specifically, excluding scales from $0.5-1\ \hMpc$ will make theories deviate by more than $1\sigma$ at scales above DES-Y3 scale cuts. 
    \item Figure \ref{fig:Mscontrib} shows that galaxy clusters contribute significantly to cosmic shears, suggesting that constraining baryonic feedback with galaxy clusters can provide constraints on the matter distribution at scales relevant to cosmic shear analyses. 
    \item Employing the DMB halo model described in section \ref{sec:theory}, we break down the matter clustering into contributions from matter residing in different environments. We calculate cosmic shear modulations, assuming that baryonic feedback only affects SZ clusters, and compare this with the total cosmic shear modulations. We find that although figures \ref{fig:rcontrib} and \ref{fig:Mcontrib} show that matter in clusters dominates matter clustering at scales relevant to cosmic shear, our calculation suggests that using SZ clusters to constrain total matter modulations still requires extrapolation of the model to lower halo masses.
    \item We test the accuracy of this extrapolation on two hydrodynamic simulations \textsc{Magneticum} and \textsc{IllustrisTNG}. In figure \ref{fig:e2e}, we show that using SZ clusters can precisely and accurately constrain the modulation of the matter distribution. Figure \ref{fig:e2e_shear} further shows that this level of precision and accuracy is subdominant to measurement errors in DES-Y3 and LSST-Y1.
    \item We perform Fisher analyses to quantify the effect of the DMB misspecification and the DMB uncertainty on the $S_8$ constraints from cosmic shear analyses in DES-Y3, LSST-Y1, LSST-Y6, and Roman. In DES-Y3 and LSST-Y1, we find a systematic bias on $S_8$ less than $0.3\sigma$ with modestly increased uncertainties $<0.08\sigma$ relative to the hypothetical perfect model of baryons (table \ref{tab:param}).
\end{itemize}

Finally, as we point out in figure \ref{fig:Mcontrib}, using SZ clusters alone to constrain total matter clustering will inevitably require extrapolations of the models. The combination of clusters from DES \citep{2016ApJS..224....1R} and the Dark Energy Spectroscopic Instrument \citep{2021ApJ...923..154T}, whose mass is calibrated with weak lensing survey is expected to constrain baryonic feedback further \citep[see e.g.][]{2020PhRvD.101d3525P}.  However, interpreting the measurements of the mean mass and the mean tSZ signal given the mass tracer of those samples, such as richness, requires understanding the selection function of those cluster finders, which is sensitive to galaxies' spatial, luminosity, and color distribution \citep{cardinal}.  Combining our DMB halo model and recent efforts on forward modeling group and cluster selection in dark matter-only simulations populated with galaxies \citep[][]{2023arXiv231003944S} will offer a powerful way for modeling the stacked tSZ signal for groups and clusters.

\appendix
\section{Modulation of matter clustering in hydrodynamic simulations} 
\label{app:hydro}
\begin{figure*}
\centering
    \includegraphics[width =0.95\textwidth]{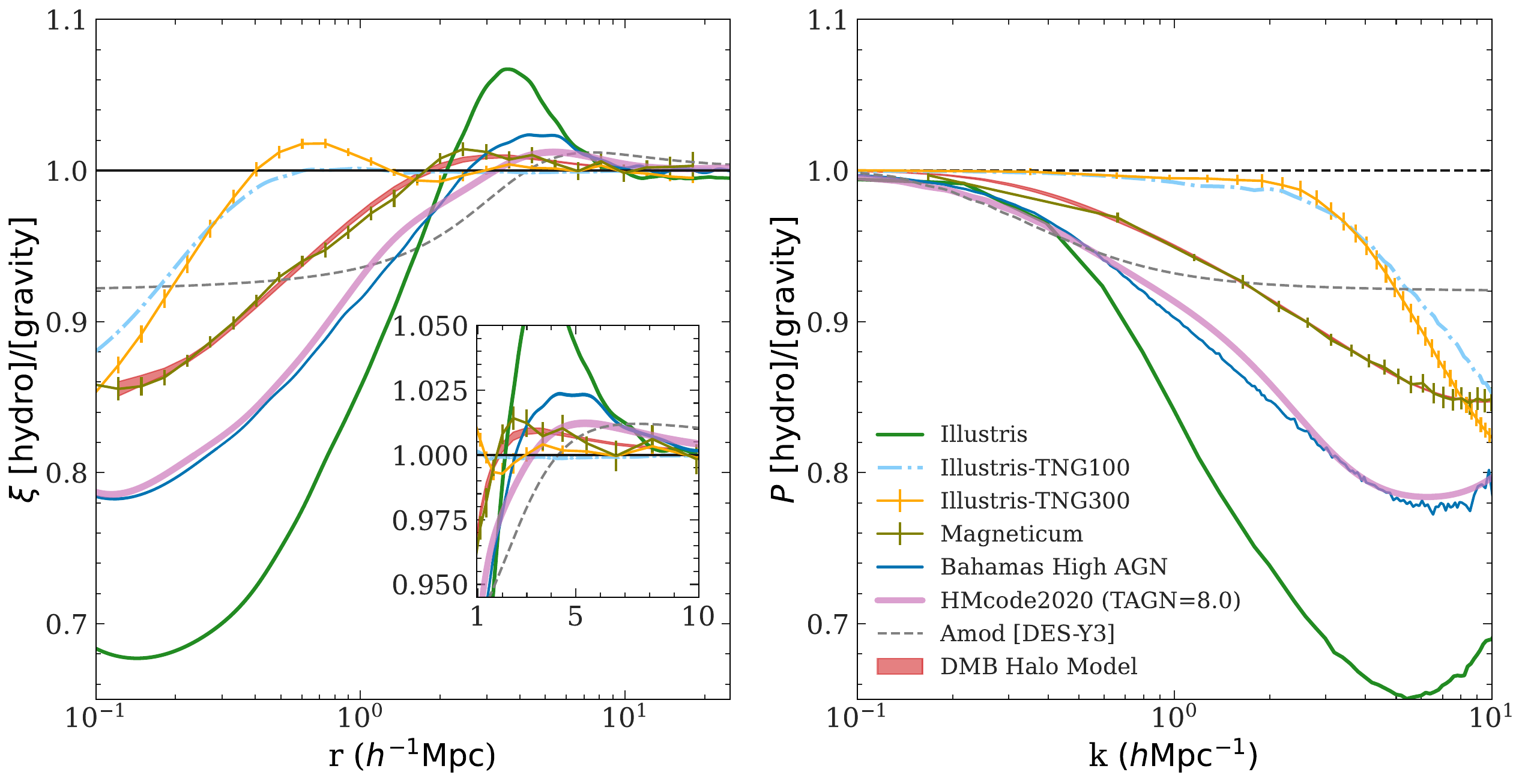}
    \caption{Modulation of the matter correlation function (left)/power spectrum (right) due to baryons at $z=0$ for different hydrodynamic simulations: Illustris (\textit{green line}) \citep{2014Natur.509..177V}, IllustrisTNG-100 (\textit{blue dot dashed line}) \citep{2019ComAC...6....2N}, IllustrisTNG-300 (\textit{yellow line with $1\sigma$ errorbars}) \citep{2019ComAC...6....2N}, Bahamas High AGN (\textit{dark blue line}) \citep{2017MNRAS.465.2936M}, Magneticum (\textit{olive line with $1\sigma$ errorbars}) \citep{2014MNRAS.442.2304H}. As a comparison, we show three empirical models: HMcode2020 \citep{hmcode2020}, Amod \citep{Amod}, and the DMB halo model presented in section \ref{sec:1h}. %
    }
    \label{fig:hydro}
\end{figure*}

As our result focuses on the matter correlation function (in configuration space), while most of the studies show modulation of matter clustering in the matter power spectrum, figure \ref{fig:hydro} shows a comparison of these measurements from several different hydrodynamic simulations. All hydrodynamic simulations predict two features in the modulation of the matter correlation function: (a) a suppression of matter clustering on small scales and (b) a boost in the matter correlation function in the 1--2 halo transition regime around $2~\hMpc$. The suppression is the well-known effect of baryonic feedback \citep[e.g.][]{vandaalen2011}, and the boost, which is not shown in the power spectra, is due to matter being redistributed by feedback to the halo outskirts. %
Further, we see that the modulation of matter correlation is highly localized, $< 1\%$ at $r=8~\hMpc$ in the most extreme hydrodynamic simulation considered here (green line, the original \textsc{Illustris} simulation \citep{2014Natur.509..177V}). %

As a comparison, we show three empirical models of matter clustering modulation: HMcode2020 \citep{hmcode2020},  $A_{\rm{mod}}$ \citep{Amod}, and the DMB halo model described in section \ref{sec:1h}. HMcode2020 is a halo model designed to fit Bahamas simulations. While we see some disagreement between HMcode2020 and the Bahamas simulation at the scales of the matter conservation boost, we note that this has minimal impact on the prediction of cosmic shear signals. This is because cosmic shear is sensitive to matter density fluctuation. While mass conservation ensures the same amount of matter in the boost and the dip, matter in the boost spreads in a much larger volume, making a smaller change in matter density. To corroborate this argument, we further calculate the differences in $\xi_-$, and $\xi_{+}$ with and without the bump and find $<1\%$ changes. $A_{\rm{mod}}$ modulates matter the power spectrum as a difference between non-linear matter clustering and linear theory predictions. Here, we show the $A_{\rm{mod}}$ value constrained from DES-Y3 data with DES-Y3 scale cuts \citep{Amod}. While the $A_{\rm{mod}}$ model captures some of the baryonic features in the matter power spectrum predicted by hydrodynamic simulations at $k<0.6\ h \rm{Mpc}^{-1}$, it deviates from those predictions in matter correlation functions even at $r=10\ h^{-1} \rm{Mpc}$. Specifically, $A_{\rm{mod}}$ pushes the matter correlation function modulation to a much larger scale than predicted by hydrodynamic simulations. This is likely because modulation due to baryonic feedback is local in configuration space, and non-linear matter clustering happens on much larger scales. Finally, we fit the DMB halo model to the correlation and power spectrum measured in Magneticum simulation. We see that this model fits both the power spectrum and correlation function very well.

\section*{Acknowledgements}
We thank Tim Eifler, Chris Hirata, and Eduardo Rozo for illuminating discussions on various aspects of this paper. We thank Priyanka Singh and Pranjal Rajendra Singh for assistance with the Magneticum measurement. CHT received support from the United States Department of Energy, Office of High Energy Physics under Award Number DE-SC-0011726. EK is supported in part by Department of Energy grant DE-SC0020247 and the David and Lucile Packard Foundation. DHW acknowledges support from National Science Foundation grant AST 2009735. 
 We acknowledge the hospitality of the Benasque Institute
in Spain, where some of the original ideas were developed. Part of this work was performed at the Aspen Center for Physics, supported by National Science Foundation grant PHY-2210452.   This research made use of SciPy \citep{Virtanen_2020} This research made use of matplotlib, a Python library for publication quality graphics \citep{Hunter:2007} This research made use of Astropy, a community-developed core Python package for Astronomy \citep{2018AJ....156..123A, 2013A&A...558A..33A} This research made use of NumPy \citep{harris2020array}

Some of the analysis was facilitated by illustris-TNG simulation \citep{2019ComAC...6....2N} and Magneticum simulation \citep{2014MNRAS.442.2304H} and \citep{2017A&C....20...52R}. We thank the authors for making simulations publicly available. 
The TNG simulations were run with compute time granted by the Gauss Centre for Supercomputing (GCS) under Large-Scale Projects GCS-ILLU and GCS-DWAR on the GCS share of the supercomputer Hazel Hen at the High Performance Computing Center Stuttgart (HLRS). The {\it Magneticum} simulations were performed at the Leibniz-Rechenzentrum with CPU time assigned to the Project `pr86re'. 
\bibliography{sample.bib} %

\label{lastpage}
\end{document}